\documentstyle[preprint,aps,eqsecnum,floats,epsf]{revtex}

\tighten

\begin{document}

\def\beq{\begin{equation}}
\def\eeq{\end{equation}}

\def\beqn{\begin{eqnarray}}
\def\eeqn{\end{eqnarray}}
\def\nn{\nonumber\\}

\def\vec#1{{\mbox{\boldmath $\bf #1$}}}

\def\qe{\vec q\cdot\e}
\def\qk{\vec k \cdot\vec q}
\def\e{\vec\epsilon}

\def\Sq{\vec\sigma\cdot\vec q}
\def\Sk{\vec\sigma\cdot\vec k}
\def\Se{\vec\sigma\cdot\vec \e}

\title {
The inclusive reaction $d(\gamma,\pi)NN$ in the first 
resonance region
}

\author
{
M.I. Levchuk$^a$  \thanks{E-mail: levchuk@dragon.bas-net.by},
M. Schumacher$^b$ \thanks{E-mail: schumacher@physik2.uni-goettingen.de}
and 
F. Wissmann$^b$   \thanks{E-mail: fwissma@gwdg.de}
}

\address
{ ~
\\ $^a$
   B.I. Stepanov Institute of Physics,
   Belarus National Academy of Sciences,
\\ F. Scaryna prospect 70, 220072 Minsk, Belarus
}

\address
{ ~
\\ $^b$
   II Physikalisches Institut der Universit\"at 
   G\"ottingen, \\ Bunsenstra\ss e 7-9, D-37073 G\"ottingen, 
Germany }

\maketitle

\begin{abstract}
Inclusive single pion photoproduction on the deuteron is studied 
in the first resonance region.  The calculation is based on the use of the 
diagrammatic approach.  Pole diagrams and one-loop diagrams with $NN$ 
rescattering in the final state are taken into account. The 
elementary operator for pion photoproduction from the  nucleon is 
taken in on-shell form and calculated  using the SAID 
and MAID multipole analyses.  
Our predictions for total and 
differential cross section show good agreement with the available
experimental data. 
Invoking some information on the reactions $\gamma 
d\to\pi^0 d$ and $\gamma d\to np$ we predict the total 
photoabsorption cross section for deuterium.  We find that our 
results overestimate the experimental  data in the center of the 
$\Delta$-peak ($\sim 320$ MeV) by about 10\%.

\end{abstract}


\section{Introduction}
\label{intr}

Recently,  comprehensive measurements of total and differential cross 
sections of inclusive, coherent and incoherent 
$\pi^0$-photoproduction from the deuteron in the energy region from 
200 to 792 MeV were carried out at MAMI \cite{krusche99}. It was 
found that the coherent data are in good agreement with theoretical 
predictions.  However, in the case of the incoherent cross sections 
the situation is much less satisfactory. The theoretical predictions 
from Refs.~\cite{laget81,schmidt96} in the $\Delta$-region are 
significantly above the data. It is evident that the model of 
Ref.~\cite{schmidt96} can hardly provide a reasonable description of 
the data on pion photoproduction from the deuteron since it takes 
into account the pole diagrams only. It is known that the effect of 
nucleon-nucleon final state interaction (FSI) is extremely important 
in incoherent photoproduction especially for small pion angles (see 
Refs.~\cite{laget81,lps96,lsw00}).
  
Although FSI was incorporated in the model of Ref.~\cite{laget81}, 
it nevertheless failed to reproduce the data. A possible 
reason for this may be that Laget used in his 
calculations of the $\gamma d\to \pi^0 np$ process the well-known 
Blomqvist-Laget (BL) parametrization \cite{BlLag77} of the pion 
photoproduction amplitude on the nucleon. This parametrization 
gives a satisfactory fit to the amplitude for charged pion 
photoproduction. But it is not able  to  describe 
$(\gamma,\pi^0)$ production from the proton. Since data on 
$(\gamma,\pi^0)$ production from the neutron are absent there 
is no possibility to check the reliability of the BL
model in the description of this channel. An attempt to remedy this 
defect in Ref.~\cite{sabutis83} led to a $\pi^0$-photoproduction 
amplitude which is not very suitable for the use in nuclear 
calculations.

In our previous analyses of $\pi^0$-photoproduction from the deuteron
\cite{lps96,lsw00} we also used the BL operator. However, 
in those cases it was quite justified due to the following reasons. 
In Ref.~\cite{lps96} we  studied the incoherent reaction 
$d(\gamma,\pi^0 n)p$ in the 
$\Delta$-region in the neutron quasi-free kinematics. 
Our main purpose was to estimate the relative contributions of the 
neutron pole diagram and background effects due to FSI and $\pi N$ 
rescattering. Since all these ingredients depend on the same pion 
photoproduction amplitude, the relative contributions are therefore not 
very sensitive to its magnitude. In Ref.~\cite{lsw00} the 
reaction  was considered in the threshold region. There, only 
the charged channels are of importance because 
of a big $\pi^{\pm} N$ rescattering effect and the use of 
the BL operator is certainly possible.

In this article we present a new computation of
the inclusive reaction $d(\gamma,\pi^0) np$ in the  first 
resonance region.  
The main difference between the present 
calculation and the one of Ref.~\cite{laget81} is, that
a more realistic version of the elementary pion photoproduction 
operator is used. 
It is taken in the standard CGLN form \cite{CGLN} with four 
partial amplitudes $F_i(\omega,\Theta_\pi)$ calculated with the use 
of the SAID \cite{arnd96} and MAID \cite{drechsel99} multipole 
analyses
\footnote { An analogous method was 
used in Ref.~\cite{levc94} when considering the reaction 
$d(\gamma,\gamma 'n) p$.  Needed in practical calculations, the 
nucleon Compton scattering operator was taken in that work in 
on-shell form with partial amplitudes obtained in the framework of 
dispersion approach \cite{lvov79} (see also Ref.~\cite{lvov97}).}. 
Since a generalization of the approach into charged pion 
photoproduction is straightforward we take the opportunity 
to consider all possible inclusive channels $d(\gamma,\pi)NN$
restricting ourselves, as in Ref.~\cite{lps96}, to the first 
resonance region. 

The paper is organized as follows.
In Sect.\ \ref{kinema}, kinematical relations used within
calculations are briefly reviewed. A description of the theoretical 
model and its ingredients is given in Sect.\ \ref{theory}. In Sect.\ 
\ref{results}, we compare our results with all available 
experimental data.  In Appendix \ref{append2} an extension of the 
non-relativistic approximations of the Bonn OBE potential is described,
in order to make them applicable to $nn$ and $pp$ scattering.
Kinematical relations between the variables in the 
so-called photon-nucleon c.m.  frame and the ones in the $\gamma d$ 
c.m.  frame are given in Appendix~\ref{append1}.

\section{Kinematics}
\label{kinema}

Let us denote by $k=(k^0,\vec k),~p_d=(\varepsilon_d,\vec 
p_d),~q=(\varepsilon_\pi,\vec q),~p_1=(\varepsilon_1,\vec p_1)$ and 
$p_2=(\varepsilon_2,\vec p_2)$ the 4-momenta of the initial photon 
and deuteron, the final pion and nucleons, respectively.
A symbol $E_\gamma$ is reserved for the lab photon energy 
($k^0_{lab}=E_\gamma$) and a symbol $\omega$ will be used for 
the photon energy in the $\gamma d$ c.m.  frame:  
$k^0_{cm}=\omega=E_\gamma M/W_{\gamma d}$ with $W_{\gamma 
d}=\sqrt{M^2+2ME_\gamma}$ and $M$ being the deuteron mass. 

It is convenient to take as  independent kinematical variables the 
photon energy and pion momentum $\vec q$ in the used frame of 
reference (generally, the lab or c.m.  frame) and the angles 
$\Theta_{\vec P}$ and $\phi_{\vec P}$ of one of the nucleons in the 
c.m. frame of the final $N_1N_2$ pair.  Using the equality
\beqn 
\label{P_cm}
W_{NN}=2\varepsilon_P=2\sqrt{{\vec  P}^2+m^2}=\sqrt{(k+p_d-q)^2},
\eeqn 
where $m$ is the averaged mass of the final nucleons, one can find 
the momentum $\vec P$. After boosting the momenta $\vec P$ and 
$-\vec P$ with the velocity $(\vec k +\vec p_d-\vec 
q)/(k^0+\varepsilon_d -\varepsilon_\pi)$ the momenta of the outgoing 
nucleons are obtained and, therefore, the kinematics is totally 
determined.

The differential cross section is given by
\beqn
\label{dcs4}
\frac {d\sigma }{d\vec q d\Omega_{\vec  P}}=
\frac 1{(2\pi)^5} 
\frac {m^2\varepsilon_d |{\bf P}|}
      {8k\cdot p_d~\varepsilon_\pi \varepsilon_P}
~~\frac 16
\sum _{m_2m_1\lambda m_d} 
| \langle m_2m_1| T| \lambda m_d \rangle |^2,
\eeqn
where $m_2$, $m_1$, $\lambda$, and $m_d$ are spin states of 
the two nucleons, photon, and deuteron, respectively.
To obtain the inclusive differential cross section $d\sigma 
/d\Omega_\pi$, the r.h.s. of Eq.\ (\ref{dcs4}) has to be integrated 
over the value of the pion momentum $q=|\vec q|$ and the solid angle
$\Omega _{\vec P}$:
\beqn
\label{dcs}
\frac {d\sigma}{d\Omega_\pi}= \int \limits_0^{q^{max}}
q^2dqd\Omega_{\vec  P}\frac {d\sigma }{d\vec q d\Omega_{\vec  P}}.
\eeqn
An extra factor of 1/2 must be included in the r.h.s. of Eq. 
(\ref{dcs}) in case of charged pion photoproduction.
The maximum value $q^{max}$ can be found from Eq.~(\ref{P_cm}) at 
$W_{NN}=2m$. In the c.m. frame it is given by
\beqn 
\label{q_cm}
q^{max}=\frac 1{2W_{\gamma d}} 
\sqrt {[W^2_{\gamma d}-(2m+\mu)^2]
       [W^2_{\gamma d}-(2m-\mu)^2] }, 
\eeqn 
where $\mu$ is the pion mass.  
In the lab frame one has
\beqn 
\label{q_lab}
q^{max}=\frac 1b \left[ aE_\gamma 
       z+(E_\gamma+M)\sqrt{a^2-b\mu^2}\right] ,
\eeqn 
where $a=(W_{\gamma d}^2-4m^2+\mu^2)/2$ and  
       $b=(E_\gamma+M)^2-E_\gamma^2z^2$ with $z=\cos{\Theta_\pi}$.

\section{The theoretical model for inclusive pion photoproduction on 
the deuteron} 
\label{theory}

As in our previous papers on $\pi^0$-photoproduction from the 
deuteron \cite{lps96,lsw00} we will exploit the diagrammatic approach 
to calculate the amplitude $\langle m_2m_1| T| \lambda m_d \rangle $.
However, we reduce the set of diagrams under consideration. 
For example, in Ref.~\cite{lsw00} working in the threshold region 
we were forced to take into account a two loop diagram which includes 
simultaneously $np$ and $\pi N$ interactions. Such a diagram is of 
importance at threshold energies since it involves a block with 
charged pion photoproduction from the nucleon. With 
increasing photon energy this diagram becomes less 
important (see Ref.~\cite{lsw00}). Above 200 MeV it can safely be 
disregarded.  It is known (see Refs.~\cite{laget78,laget81}) that 
there are kinematical regions where a one loop diagram with $\pi N$ 
rescattering  noticeable contributes to the amplitude. But this 
rather concerns the exclusive process $\gamma d\to \pi NN$. We have 
checked that $\pi N$ rescattering changes the final results  in the 
first resonance region only by a few percent.

\begin{figure}[p]
\epsfxsize=0.85\textwidth
\centerline{\epsfbox[0 540 550 770]{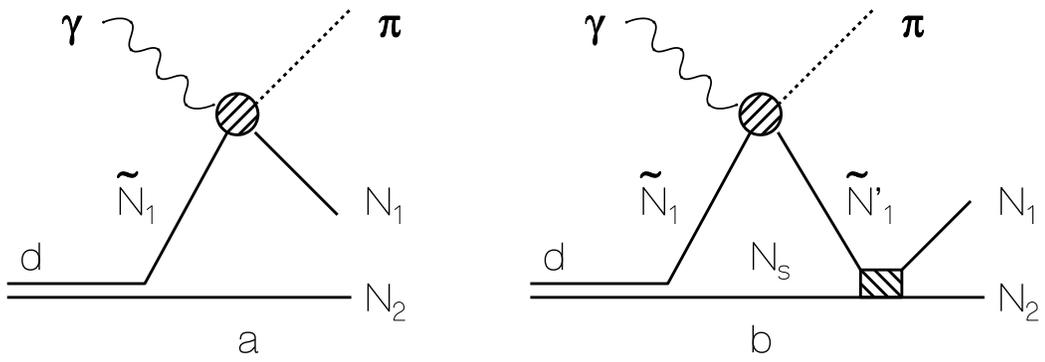}}
\caption{Diagrams considered in this work. Two other diagrams
with the permutation $1\leftrightarrow 2$ are assumed.}
\label{fig1}
\end{figure}

As a result, we retain in our calculations the two diagrams shown in 
Fig.~\ref{fig1}. The pole diagram \ref{fig1}$a$ must be considered 
since at the integrations in Eq. (\ref{dcs}) there are the 
kinematical regions where one of nucleons (or both simultaneously)  
has a small momentum, the so-called quasi-free regions.  
These lead to peaks in the exclusive cross sections.  The inclusive 
cross section from the pole diagrams is mainly saturated in these 
peaks.  When at the integration  mentioned above $q$ is 
approaching $q^{max}$ and, therefore, $W_{NN}\to 2m$, the 
relative momenta ($\sim|{\bf P}|$) of the outgoing nucleons become 
small.  Now there are peaks in the exclusive cross sections due to 
strong $NN$ interaction in the $s$-waves (see, e.g., 
Refs.~\cite{laget78,laget81,levc94}) which manifest themselves in a 
big contribution of diagram \ref{fig1}$b$ to the inclusive cross 
section.  This effect is expected to be most pronounced at small 
pion angles since 
in this case the kinematics permits both for the deuteron wave  
function (DWF) and $NN$ scattering amplitude to work simultaneously  
in low momentum regime.  The above mentioned smallness of the $\pi N$ 
rescattering effects can be explained also by the fact that the 
$s$-wave $\pi N$ scattering lengths are about two orders smaller than 
those for $NN$ scattering.

Let us now write out the matrix elements corresponding 
to the diagrams in Fig.~\ref{fig1} (see also 
Refs.~\cite{laget78,laget81,lsw00}).  One has for the pole diagram 
\ref{fig1}{\it a}
\beq
\label{pole}
\langle m_2m_1| T^a(\vec k,\vec q,\vec p_2) 
| \lambda m_d\rangle =\sum _{\tilde m_1}
\Psi ^{m_d}_{m_2\tilde m_1}\left({\vec p}_2-\frac {\vec p_d}2 
                               \right)
\langle m_1| 
T_{\gamma {\tilde N_1}\to \pi N_1}(\vec k_{\pi N_1},\vec q_{\pi N_1})
| \lambda \tilde m_1 \rangle, 
\eeq 
where $\Psi ^{m_d}_{m_2\tilde m_1}({\vec 
p}_2-\vec p_d/2)$ is DWF and 
$
\langle m_1| 
T_{\gamma {\tilde N_1}\to \pi N_1}(\vec k_{\pi N_1},\vec q_{\pi N_1})
| \lambda \tilde m_1 \rangle
$ is the amplitude of the elementary process 
$\gamma  N\to \pi N$. The amplitude depends on photon ($\vec k_{\pi 
N_1}$) and pion ($\vec q_{\pi N_1}$) momenta taken in the c.m. frame 
of the $\pi N_1$ pair. These momenta can be obtained from the 
corresponding momenta in the used frame of reference through a 
boost with the velocity $(\vec p_2-\vec k-\vec 
p_d)/(k^0+\varepsilon_d-\varepsilon_2)$.

Of course there is one more pole diagram identical to that in  
Fig.~\ref{fig1}{\it a} but with the replacement $1\leftrightarrow 2$.  
In case of $\pi^0$-photoproduction the corresponding matrix element 
should be added to Eq.\ (\ref{pole}). For the charged channels a 
subtraction of  two matrix elements should be done.

The calculations were done using DWF for the non-relativistic  
versions of the Bonn OBE potential (OBEPR) \cite{mach87,mach89}. In 
fact, in those papers three OBEPR models where built.  A 
parametrization for one of them was given in Table~14 of 
Ref.~\cite{mach87}. Two other parametrizations were given in  
Table~A.3 of Ref.~\cite{mach89} denoted in that table by ``A'' and 
``B''.  For these three versions we shall use the notations 
``OBEPR'',``OBEPR(A)'' and ``OBEPR(B)'', respectively. 
Analytical parametrizations of the $s$- and $d$-amplitudes of DWF 
for these three models were taken from Ref.~\cite{lev95}.  We 
would like to note here that  our results are practically independent 
of the choice of the potentials so that all results below were 
obtained with the OBEPR model. 

It has been noted in Sect. \ref{intr} that in our previous papers  
\cite{lps96,lsw00} on $\pi^0$-photoproduction from the deuteron we 
used the BL operator to calculate the amplitude $\langle m_1| 
T_{\gamma {\tilde N_1}\to \pi N_1} | \lambda \tilde m_1 \rangle 
$. In the present paper the latter was obtained with the use of 
multipole analyses.  
The CGLN operator has the following form:
\beqn
\label{CGLN}
T_{\gamma N\to \pi N}=\frac {4\pi W_{\gamma N}}m 
\left[
i \Se_\lambda ~F_1 + \vec q \cdot (\vec k \times \e_\lambda)~ F_2 + 
i\Sk ~\qe_\lambda ~F_3 + i\Sq ~\qe_\lambda ~F_4 
\right]. 
\eeqn
It is written in the $\pi N$ c.m. frame so that all vectors in Eq. 
(\ref{CGLN}) should be taken in this frame. Of course, the 
partial amplitudes $F_i(\omega,\Theta_\pi)~(i=1-4)$ are also 
functions of the photon energy $\omega$ and pion angle $\Theta_\pi$ 
in the same frame. Therefore,  in practical calculations, in particular
at the numerical integration in Eq.~(\ref{dcs}) (and in Eq.~(\ref{np-resc}),
see below) one has 
to make Lorenz transformations to this frame for every grid point. 
But such a procedure is 
of no principal difficulty and does not require time-consuming 
computations. We do not give here explicit expressions for the amplitudes 
$F_i(\omega,\Theta_\pi)$ through electric and magnetic multipoles and 
the  derivatives of the Legendre polynomials \cite{CGLN} since they 
are very well-known. 
The multipoles are taken from the SAID 
\cite{arnd96} and MAID \cite{drechsel99} analyses. If not stated 
otherwise all results below have been obtained with the SAID 
multipoles.

The matrix element corresponding to diagram \ref{fig1}{\it b} is
\beqn
\lefteqn{
\langle m_2m_1| T^b(\vec k,\vec q,\vec p_2) 
| \lambda m_d\rangle = } \nonumber \\
&&m\int \frac {d^3{\vec p}_s}{(2\pi)^3} 
\sum _{\tilde m_2'\tilde m_1'}
\frac 
{\langle{\vec p}_{out},m_2m_1| T_{NN}| {\vec p}_{in},
 \tilde m_2'\tilde m_1'\rangle\langle
 \tilde m_2'\tilde m_1'| T^a(\vec k,\vec q,\vec p_s) 
| \lambda m_d\rangle}
{p^2_{in}-p^2_{out}-i0},
\label{np-resc}
\eeqn
where ${\vec p}_{out}=({\vec p}_2-{\vec p}_1)/2$ and ${\vec 
p}_{in}={\vec p}_s+({\vec q}-{\vec p_d}-{\vec k})/2$ are the relative 
momenta  of the $N_1N_2$ pair after and before scattering, 
respectively, and $\langle{\vec p}_{out},m_2m_1| T_{NN} | {\vec 
p}_{in}, \tilde m_2'\tilde m_1'\rangle$  is the half off-shell $NN$ 
scattering amplitude. We will not discuss here details of the
computations of the amplitude (\ref{np-resc}) because they are given 
in Ref.~\cite{lps96}. Note that all partial waves with the 
total angular momentum $J=0$ and 1 were retained in the $NN$ 
scattering amplitude. In fact, however, only two waves, $^1S_0$ and 
$^3S_1$, are of importance when the inclusive channels are 
considered. All other waves give a few percent contribution to the 
cross section.

The same OBEPR models of $NN$ interaction were used when  solving the 
Lippmann-Schwinger equation for the $NN$ scattering amplitude needed 
for the calculations of diagram \ref{fig1}{\it b}. It must be 
noted, however, that those models are valid for  $np$ interaction 
only.  Therefore, they should be modified in such a way that they 
may be applicable to $nn$ and $pp$ interactions as well.  We follow 
a procedure for such a modification proposed in 
Ref.~\cite{haidenb89}.  It is described in some detail in Appendix 
\ref{append2}. 

As in our previous papers \cite{lps96,lsw00}, 
all summations over polarizations of the particles in Eqs.  
(\ref{pole}) and (\ref{np-resc}) as well as the three-dimensional  
integrations in Eq. (\ref{np-resc}) have been carried out 
numerically. The number of chosen nodes at this integration and that 
in Eq. (\ref{dcs}) was taken to be sufficient for prediction of the 
differential cross section with the numerical accuracy better than
2\%.

\section{Results and Discussion}
\label{results}

We begin our discussion with the results for the $d(\gamma,\pi^0)np$ 
channel. 
In Fig.~\ref{fig2}, the predicted differential cross 
sections of this reaction are shown at energies between 208 and 456 
MeV together with experimental results from Ref.~\cite{krusche99} 
\footnote
{
In Ref.~\cite{krusche99} the differential cross sections 
are given in the so-called ``photon-nucleon c.m. frame''. Relations 
needed to transform the cross sections and angles from  the $\gamma 
d$ c.m. frame to the frame mentioned are presented in Appendix\ 
\ref{append1}.
}.
One can see one more confirmation of a prediction of 
Refs.~\cite{laget81,lps96} that the effect of $np$ final state 
interaction should lead to a reduction of the cross section 
and this reduction is the stronger the smaller the pion angles are. 
This effect is mainly attributed to the strong repulsive $np$ 
interaction in the $^3S_1$ wave.

Without FSI the model totally fails to reproduce the data.
After including FSI one has a quite reasonable description of the 
data at all energies except for $E_\gamma =208$ MeV in the
backward direction. Only the points corresponding  to
$\Theta^{*N}_\pi=110^\circ$ at energies from 285 to 362 MeV are 
noticeably below the curves. But it is difficult to draw smooth 
curves through the data if these points are included. 

In the same figure we compare our results with those from 
Refs.~\cite{laget81,schmidt96}. Since in Ref.~\cite{schmidt96} FSI 
was disregarded one could expect the dotted curves to be close to the 
predictions of that work. In fact, one has some deviation which is 
reduced when the energy approaches the $\Delta$-position.
A reason for this deviation may be due to the use of different 
pion photoproduction operators since the only remaining 
ingredient of the models, namely the deuteron wave functions, 
are very similar for all modern $NN$ potentials. Indeed, a comparison 
of predictions for the total cross sections of the reaction $\gamma 
p\to\pi^0 p$ given in Fig.~2 of Ref.~\cite{schmidt96} with those 
calculated with the SAID (and MAID) multipoles shows that these 
former go above the latter.

An analogous reason seems to be responsible for the disagreement 
between our full calculation and the predictions from  
Ref.~\cite{laget81} where FSI was taken into account. As already 
mentioned in Sect.\ \ref{intr}, the BL operator is not good for 
the description of neutral pion photoproduction.  The deviation is 
clearly seen at $\Theta^{*N}_\pi\ge 90^\circ$ above 300 MeV.

To illustrate, we show at 324 MeV the contribution from the pole 
diagram with $\pi^0$-photoproduction from the proton. A constructive 
effect of two mechanisms with quasi-free pion  photoproduction 
on separate nucleons is obvious.  One can see that at backward 
pion angles the total contribution of the pole diagrams is practically 
equal to the direct sum of the contributions of each diagram so that the 
interference term is very small. The 
reason for this is that at backward angles the kinematics does not 
allow both nucleons to have simultaneously small momenta and, 
therefore, both diagrams cannot work in the quasi-free regime.

After integrating Eq. (\ref{dcs}) over the solid pion angle one obtains 
the total cross section for a given channel. In Fig.~\ref{fig3},
the total cross section for $\pi^0$-photoproduction is shown. As in 
the case of the differential cross section, one can see that without 
FSI the model clearly overestimates the data. After inclusion of FSI 
one has good agreement with the data. Only in the center of the peak 
our model overestimates the measured cross sections by about 
25 $\mu b$.  Note that the inclusion of FSI does not shift the position
of this peak. This fact has a simple explanation. Indeed, diagram 
$b$ in Fig.~\ref{fig1} contains the same $\gamma N\to\pi N$ 
amplitude as the pole diagram $1a$. Taking into account that the 
integral over the loop is saturated at small momenta due to the 
deuteron wave function, the boosts mentioned in Sect.\ \ref{theory} 
lead only to small shifts in energies. Therefore, the $\gamma 
N\to\pi N$ vertex works practically in the same energy regime as it 
does in the pole diagrams. 

It is clear that the calculation of Ref.~\cite{schmidt96} totally 
fails to describe the data, as can be expected from the previous 
discussion. Although FSI was taken into account in  
Ref.~\cite{laget81}, the predictions from that work give still too 
high cross sections.

We begin the discussion of charged pion photoproduction with 
the $d(\gamma,\pi^-)pp$ channel. There is one experimental article 
\cite{benz73} which supplies us with a lot of data points in 
the energy region from 0.2 to 2.0 GeV. However, we will discuss the 
first resonance region only. In Fig.~\ref{fig4} the 
predicted differential cross  sections are shown at energies 
ranging from 210 to 540 MeV.  The dotted curves which correspond to 
the contribution of one pole diagram, reproduce the behaviour of the 
angular dependence for the differential cross section of the 
elementary reaction $\gamma n\to\pi^-p$. In particular, at energies 
above 400 MeV the strong forward peak due to the contribution of the 
pion exchange in the $t$-channel is clearly seen. After inclusion of 
the second pole diagram one has a drastic reduction of the cross 
section at forward pion angles, exhibiting the total difference from 
the case of neutral pion photoproduction and showing how the Pauli 
principle manifests itself. One can again see that  at backward angles 
the cross section from two pole diagrams is practically equal to twice 
the cross section from one diagram.

The effect of FSI in the case of the charged channels is expected to 
be quite different in comparison with the neutral one. Since only 
$s$-wave $NN$ interaction is of importance for the inclusive 
$d(\gamma,\pi)NN$ reaction, FSI shows up in the charged 
channels through attractive interaction in the $^1S_0$ wave. 
This interaction is again significant at small pion angles but 
has to lead to an increase of the differential cross section. 
A confirmation of our anticipations is seen in Fig.~\ref{fig4}. 
A very interesting observation is that above 400 
MeV the cross section again becomes to be strongly forward peaked 
but now due to FSI. After inclusion of FSI we obtain good  
agreement with the data from Ref.~\cite{benz73}. 

At 350 MeV we compare our results with those from 
Ref.~\cite{laget81}. One can see that there is agreement in the 
shapes of the angular distributions but we predict lower values for 
the cross sections.

The total cross section for $\pi^-$-photoproduction is shown in 
Fig.~\ref{fig5}. Here the contribution of FSI is noticeable smaller 
than that for $\pi^0$-production and leads to an increase of the 
cross section. We find good  agreement with the data from  
Refs.~\cite{benz73} and \cite{chief75} but the data from  
Ref.~\cite{asai90} lie above our solid curve at $E_\gamma \ge 375$ 
MeV.  At the same time a data pion from Ref.~\cite{quraan98} at 250 
MeV lies markedly below both our predictions and the data from  
Refs.~\cite{benz73} and \cite{chief75}. Theoretical predictions from 
Refs.~\cite{laget81,schmidt96} are also shown in Fig.~\ref{fig5}.  
They are very close to each other and are able to reproduce the data 
only below 250 MeV.

In  Figs.~\ref{fig6} and \ref{fig7} we give our results 
for $\pi^+$-photoproduction. All the theoretical conclusions we have 
just drawn for the case of $\pi^-$-photoproduction are
valid for the $\pi^+$ channel as well. To our knowledge 
there  are no data on this process in the first resonance region so  
that we cannot compare our predictions with experimental results.

Having results for the total cross sections in all the 
channels mentioned above one can try to make predictions for the 
total photoabsorption cross section on the deuteron in the first 
resonance region.  Of course, two more reactions contribute to it as 
well.  These are coherent $\pi^0$-photoproduction from the 
deuteron ($\gamma d\to\pi^0 d$) and deuteron photodisintegration 
($\gamma d\to np$). Predictions for the former are taken from a 
theoretical paper \cite{kamalov97} which are in good agreement with 
the data from Ref.~\cite{krusche99}. The total cross sections for 
the latter reaction were calculated making use of 
a phenomenological fit \cite{ross89} to available experimental data 
up to 440 MeV.

In Fig.~\ref{fig8} we present our results for 
the total photoabsorption cross section per nucleon for the deuteron. Good 
agreement with the data is seen excluding, however, the $\Delta$-peak 
region. In the center of the peak at about 320 MeV  we find our 
results with SAID and MAID multipoles to overestimate the 
experimental value of $(452\pm 5)~\mu b$ by 33 $\mu b$ and 48 
$\mu b$, respectively.  We have no explanation for this disagreement.

It is instructive to compare the results of the direct measurements 
of the total photoabsorption cross section for the deuteron from 
Refs.~\cite{amstrong72,maccornick96} and those which can be extracted 
from the data on the separate channels contributing to this cross 
section.
Let us try to put together all available data at 320 MeV.
From Ref.~\cite{krusche99} one has for the $d(\gamma,\pi^0)np$ and
$d(\gamma,\pi^0)d$ channels $(348\pm 27)~\mu b$ and $(124\pm 19)~\mu 
b$, respectively  (the additional normalization error of 6\% has been 
added).  The result for  the $d(\gamma,\pi^-)pp$ channel from 
Refs.~\cite{benz73,chief75} is $(249\pm 10)~\mu b$ (again the 
additional normalization error of 5\% has been added).  At 320 MeV 
the phenomenological fit \cite{ross89} gives $53~\mu b$ for the 
contribution of deuteron photodisintegration. The errors in this 
number can be safely neglected. Since there are no data on the 
$d(\gamma,\pi^+)nn$ channel in the $\Delta$-region we accept for its 
contribution our theoretical  prediction of $199~\mu b$. Collecting 
all these numbers we obtain the value of $(487\pm 17)~\mu b$ for the 
total photoabsorption cross section per nucleon for the deuteron at 
320 MeV which is in disagreement with the results from  
Refs.~\cite{amstrong72,maccornick96}.  However, there exists good 
agreement with our values of $485~\mu b$ and  $500~\mu b$ obtained 
with the SAID and MAID multipoles, respectively. 

\section{Summary}
\label{summary}

We have investigated inclusive single pion photoproduction on the 
deuteron in the first resonance region.  Unlike most previous calculations, 
although not numerous, we have exploited as the elementary  operator 
for pion photoproduction from the  nucleon the one calculated  with 
the SAID and MAID multipoles rather than the commonly used 
Blomqvist-Laget one or an operator built in Ref.~\cite{schmidt96}.  
We have found that the model involving the pole diagrams and FSI 
gives a good description of all available data both on the 
differential and total cross sections.  This description is much 
better than that in other theoretical approaches.  The only 
unsolved problem is that our predictions for the total 
photoabsorption cross section for the deuteron in the $\Delta$-peak 
region overestimate the data by about 10\%. We suppose, however, that 
there is some inconsistency in the experimental results. New 
measurements, both  inclusive single pion photoproduction on the 
deuteron in all inelastic channels and total photoabsorption cross 
section for the deuteron, would be extremely desirable and could shed 
more light on possible reasons for the above disagreement.

{\acknowledgments
We would like to thank B. Krusche and A.I. L'vov for many fruitful 
discussions. We are very grateful to S.S. Kamalov for providing 
us with the results of his calculations and to R. 
Machleidt for a computer code for the CD-Bonn potential.  This work 
was supported by Deutsche  Forschungsgemeinschaft under contract 436 
RUS 113/510.}

\newpage
\appendix

\section{Application of the Bonn OBEPR model to neutron-neutron 
and proton-proton scattering.}
\label{append2}

Used through this article the Bonn OBEPR model \cite{mach87,mach89} 
is valid, strictly speaking, for $np$ interaction only. 
Since in our consideration of charged pion photoproduction the 
amplitudes of $nn$ and $pp$ interactions are also needed we have to 
modify the potential so that it could be applicable to  these 
interactions too. To do this we will use a procedure proposed in  
Ref.~\cite{haidenb89} which consists in adding Coulomb interaction  
to the original Bonn model and making small adjustments of the 
parameters of the model  
\footnote { It would be much more consistently to deal
with a model which includes the charge dependence of nuclear forces 
and describes simultaneously all channels in $NN$ interaction. 
Such a model called 
`CD-Bonn' has recently been built by Machleidt \cite{mach00}. 
Unfortunately, all our main calculations had already been finished 
when we got to know about the new potential. Very first estimates 
made with this potential show only small variations, within less than 
1\%, of the presented results.}. 

A method to handle Coulomb interaction  in momentum space was 
proposed by  Vincent and Phatak \cite{vincent74}. We will not 
discuss it here since  it is described in full detail in that article
(see also Refs.~\cite{haidenb89,mach00,holz89}). We only 
mention that we applied the method to the $^1S_0$ partial wave. 
All other waves with $J=0$ and 1 were taken for the switched off 
Coulomb potential. It makes no sense to include the Coulomb 
modifications for the waves other than $^1S_0$, since the
contributions of these former to the differential and total cross 
sections were found to be very small.   

Coulomb interaction is mainly responsible for the 
difference of the $nn$ and $pp$ scattering lengths. But the 
difference of the $nn$ and $np$ scattering lengths is due to the 
breaking of the charge  independence of the nuclear force.
The major reason for it is the pion mass splitting. This effect 
was not taken into account in the construction of the Bonn 
potentials.  At least quantitatively, the differences above can be 
modelled by a procedure given in Ref.~\cite{haidenb89}. In that
article it was proposed to vary the coupling constants of the 
$\sigma NN$ (for the isospin $I=1$ channel) and $\delta NN$ vertices 
keeping all other parameters of the model unchanged.  These coupling 
constants were changed in such  a way that the new model with the 
switched on Coulomb potential describes the $^1S_0$ scattering length 
in the $pp$ channel.  At the same time in the $^3S_1$--$^3D_1$ 
partial wave there were the same deuteron properties as for the 
original version of the potential.  Carrying out the proposed 
procedure we obtained the following coupling constants for the OBEPR, 
OBEPR(A) and OBEPR(B) models, respectively:  7.7135, 8.6226 and 
8.7316 for $g^2_\sigma /4\pi$ ($cf.$ the former values 7.7823, 8.7171 
and 8.8322) and 2.586, 2.81 and 6.744 for $g^2_\delta /4\pi$ ($cf.$ 
the former values 2.6713, 2.742  and 6.729).  The resulting 
scattering lengths and effective ranges for three OBEPR models 
together with experimental values are given in Table~\ref{tab1}.

Solving the Lippmann-Schwinger equation with the original and 
modified versions of the OBEPR models we directly obtained the partial 
half-off shell amplitudes for $np$ and $nn$ scattering, respectively. 
For the $pp$ scattering we used a prescription from 
Ref.~\cite{kolybasov75} consisting of the following parametrization 
of the half-off-shell $^1S_0$ partial amplitude: 
\beqn
t_{off}^{^1S_0}(|\vec p_{out}|,|\vec p_{in}|) =
\frac {|\vec p_{out}|^2+\beta^2} {|\vec p_{in}|^2 +\beta^2}~
t_{on}^{^1S_0}(|\vec p_{out}|,|\vec p_{out}|), 
\eeqn
with $\beta=1.2$ fm$^{-1}$ and the on-shell amplitude 
$t_{on}^{^1S_0}(|\vec p_{out}|,|\vec p_{out}|)$ is obtained with the
use of the  Vincent and Phatak method  for the modified 
potentials with switched on Coulomb interaction.

\section{Kinematical relations between the variables in
photon-nucleon c.m. frame and the $\gamma d$ c.m. frame.} 
\label{append1}

In Ref.~\cite{krusche99} the differential cross sections for the 
reaction $d(\gamma,\pi^0)np$ are given in the so-called photon-nucleon c.m. 
frame. This latter corresponds to an assumption that all nucleons in a 
nucleus (we suppose the nucleus with mass number $A$ but not only the 
deuteron) have the same momenta $-\vec k /A$ in the $\gamma A$ c.m. 
frame. This means that in the lab frame all nucleons in 
the nucleus are at rest and, therefore, the total energy of the 
$\gamma N$ system is equal to $W_{\gamma N}=\sqrt{m^2+2mE_\gamma}$. 
The well known formulae give the photon energy and pion momentum in 
the $\gamma N$ c.m. frame:
\beqn 
\label{qcm}
\omega^\ast=\frac m{W_{\gamma N}}E_\gamma,\quad q^\ast=\frac 
1{2W_{\gamma N}} \sqrt {[W^2_{\gamma N}-(m+\mu)^2] [W^2_{\gamma 
N}-(m-\mu)^2] },
 \eeqn 
and the pion energy is 
$\varepsilon_\pi^\ast=\sqrt{{q^\ast}^2+\mu^2}$.

One can show that the photon energy and pion momentum in the frame 
where the nucleon has the momentum $-\vec k /A$ are expressed through 
the following relations:
\beqn 
\label{omega1}
\tilde \omega=\frac { W_{\gamma N}^2-m^2}
{2\sqrt{ \frac {W_{\gamma N}^2+(A-1)m^2}A} },
 \eeqn 

\beqn 
\label{q1}
\tilde q=\frac 1b_1 
\left[ 
a_1 \omega_A z+d\sqrt{a_1^2-b_1\mu^2}
\right] ,
\eeqn 
where $a_1=(d^2-m^2+\mu^2-\omega_A^2)/2$, $b_1=d^2-\omega_A^2z^2$, 
$d=\tilde \omega +\sqrt{(\tilde \omega/A)^2+m^2}$, $\omega_A=\tilde 
\omega(A-1)/A$ and $z$ the cosine of the pion angle in the $\gamma A$ 
c.m. frame. Of course, at $A=1$ one gets from Eqs.\ (\ref{omega1}) 
and (\ref{q1}):  $\tilde \omega =\omega^\ast$ and $\tilde q =q^\ast$.

Using Eqs. (\ref{qcm})--(\ref{q1}) one can express $z^{\ast 
N}= \cos{\Theta^{\ast N}_\pi}$ through $z$
\beqn 
\label{zN}
z^{\ast N}=\frac {\omega^\ast \varepsilon_\pi^\ast -
(\tilde \varepsilon_\pi-\tilde q z)\tilde \omega}
{\omega^\ast q^\ast}.
 \eeqn 
The derivative of $z^{\ast N}$ with respect to $z$ is then 
given by
\beqn 
\label{dzNz}
\frac {\partial  z^{\ast N}}{\partial z}=
\frac {\tilde \omega}{\omega^\ast q^\ast}
\left[ \left(z-\frac {\tilde q} {\tilde \varepsilon_\pi}\right)
\frac {\partial \tilde q}{\partial z}
+\tilde q\right].
 \eeqn 
Presented in Fig.~\ref{fig2} the differential cross sections and 
angles $\Theta_\pi^{*N}$ were  obtained making use of Eqs.\ 
(\ref{zN})--(\ref{dzNz}) for $A=2$.

\begin{table}
\caption 
{
Effective range parameters of the $^1S_0$ wave.
}
\begin{center}
\begin{tabular}{l|cccc}
\multicolumn{1}{l|}{} &
\multicolumn{2}{c}{$nn$ channel} & 
\multicolumn{2}{c}{$pp$ channel} \\
     &$a_s$ (fm) &$r_s$ (fm) &$a_s$ (fm) &$r_s$ (fm) \\
\hline
OBEPR      & $-17.78$    & 2.71  & $-7.82$   & 2.64     \\
OBEPR(A)   & $-17.82$    & 2.75  & $-7.82$   & 2.62     \\
OBEPR(B)   & $-17.73$    & 2.76  & $-7.82$   & 2.63     \\
Experiment & $-18.9  \pm 0.4   $\cite{howell98,gonzales99}
           & $2.75   \pm 0.11  $\cite{miller90}
           & $-7.8149\pm 0.0026$\cite{berg88}     
           & $2.790\pm 0.014   $\cite{berg88}    
\\ 
\end{tabular} 
\end{center} 
\label{tab1} 
\end{table}

\newpage
~

\newpage

\begin{figure}
\epsfxsize=0.85\textwidth
\centerline{\epsfbox[50 30 550 765]{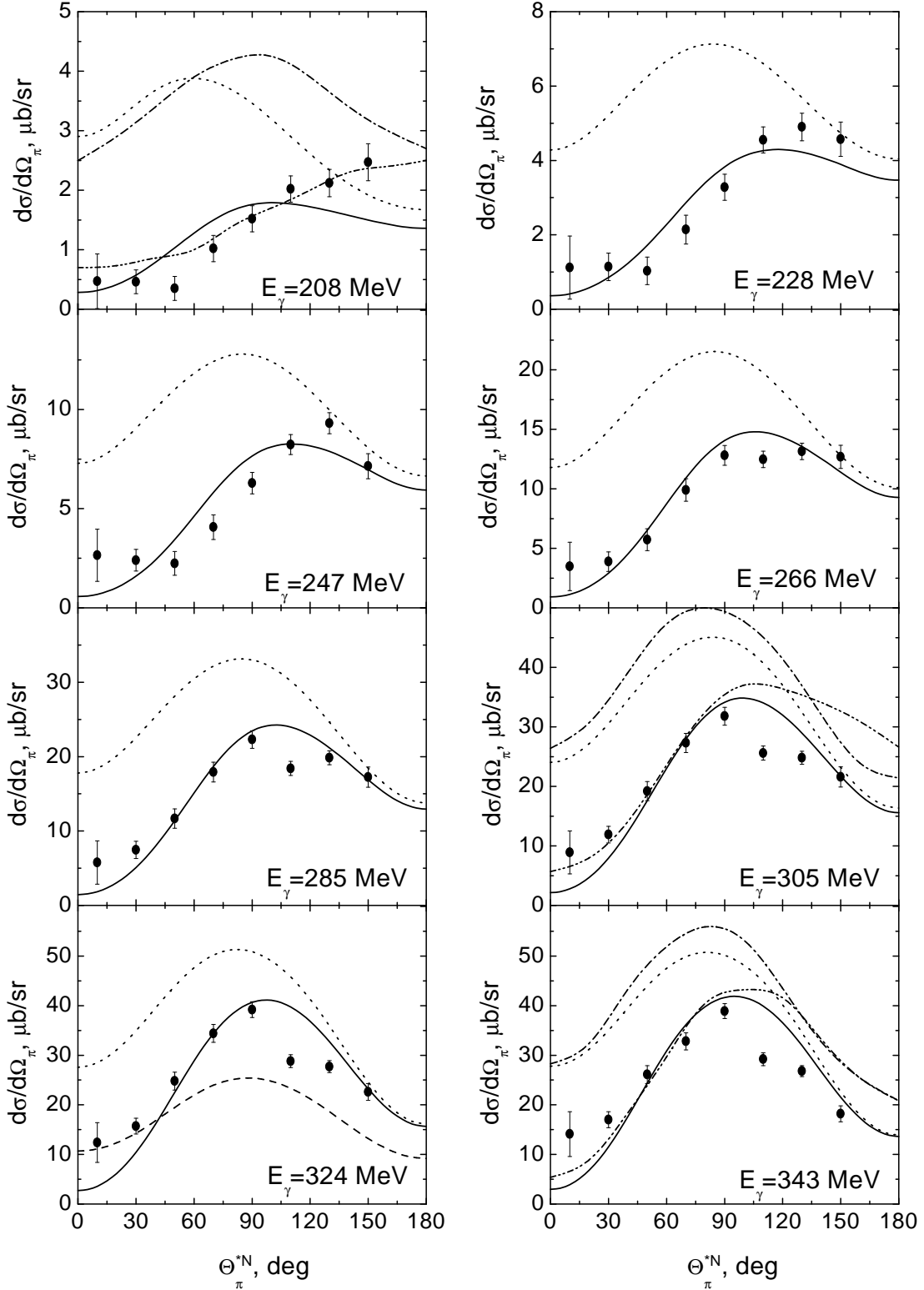}}
\caption
{
Differential cross sections for the reaction 
$d(\gamma,\pi^0)np$ in the photon-nucleon c.m. frame. The 
dotted (full) curves are our predictions without (with) FSI. 
At 324 MeV the contribution of the pole diagram with pion production 
from the proton is shown in dashed curve.  The dash-double-dotted and 
dash-dotted curves are results of Refs.~{\protect \cite{laget81}} 
and {\protect \cite{schmidt96}}, respectively, borrowed from  
Ref.~{\protect \cite{krusche99}}.  Data are from Ref.~{\protect 
\cite{krusche99}}.  
} 
\label{fig2} 
\end{figure}

\setcounter{figure}{1}
\begin{figure}
\epsfxsize=0.85\textwidth
\centerline{\epsfbox[50 200 550 810]{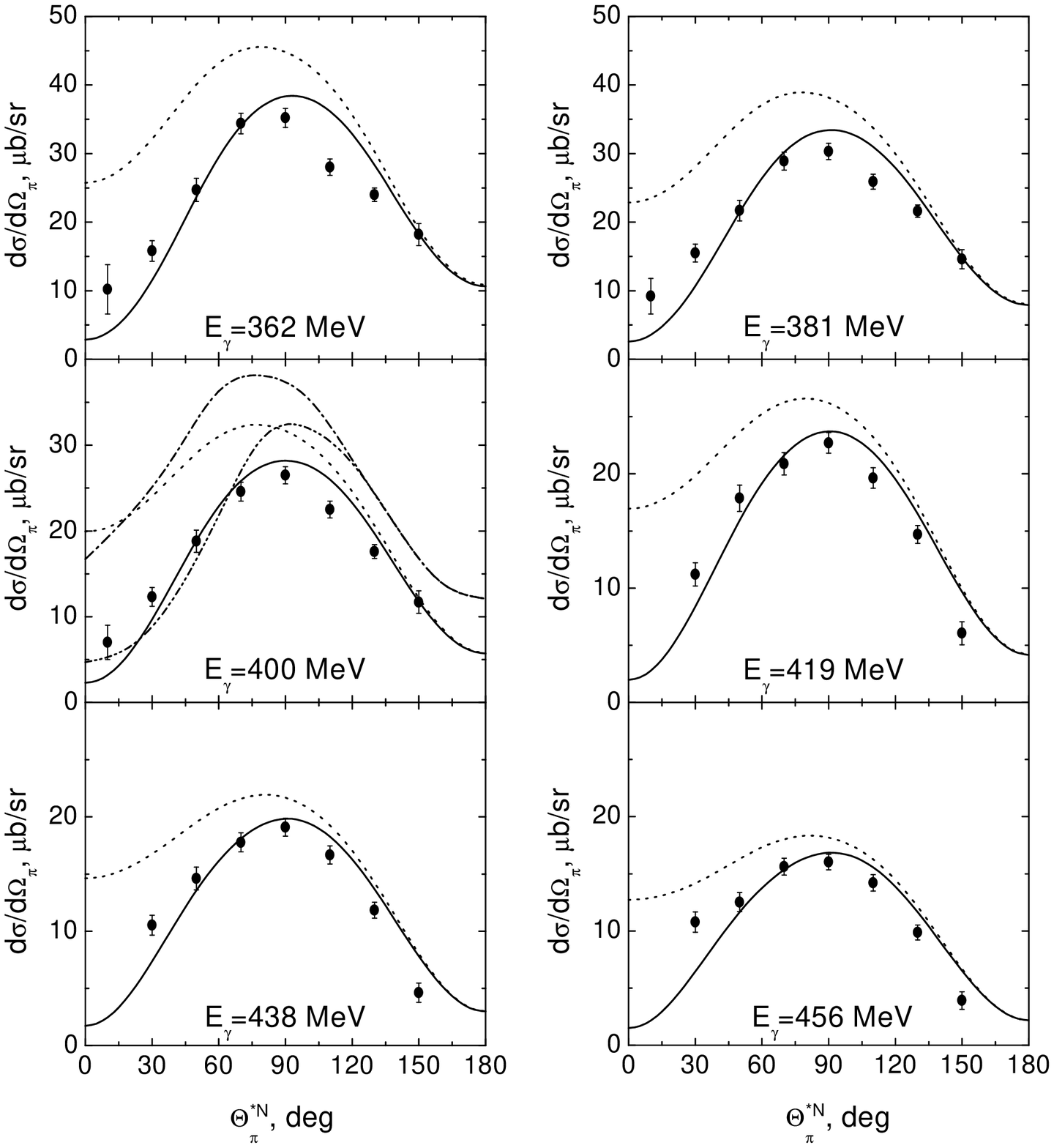}}
\caption{ Continued.
}
\label{fig2_2}
\end{figure}

\begin{figure}
\epsfxsize=0.8\textwidth
\centerline{\epsfbox[40 330 450 810]{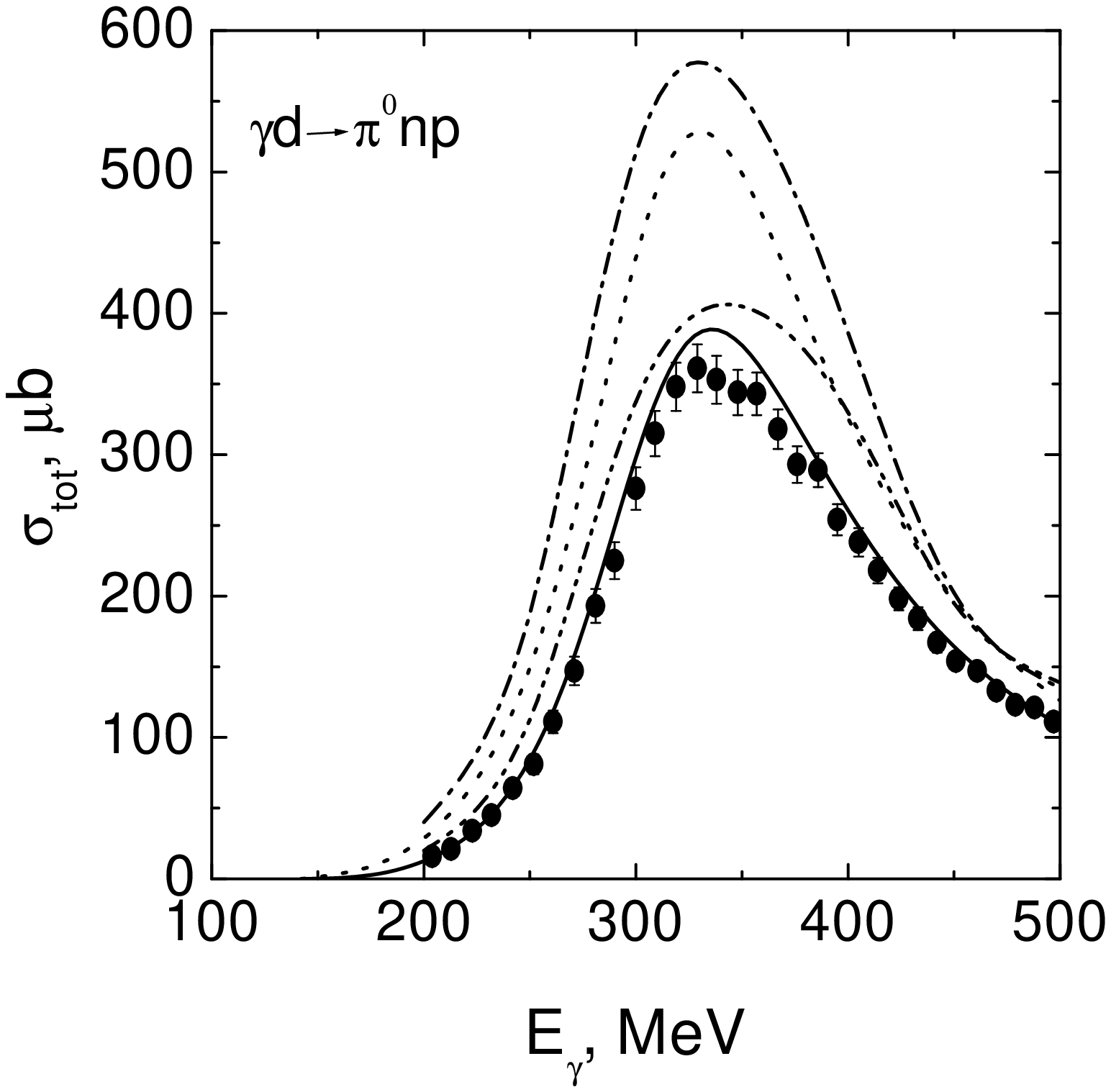}}
\caption
{ 
Total cross section for the reaction 
$d(\gamma,\pi^0)np$. Meaning of the curves as in Fig.~\ref{fig2}.
Data are from Ref.~{\protect \cite{krusche99}}.
}
\label{fig3}
\end{figure}

\begin{figure}
\epsfxsize=0.85\textwidth
\centerline{\epsfbox[50 30 550 750]{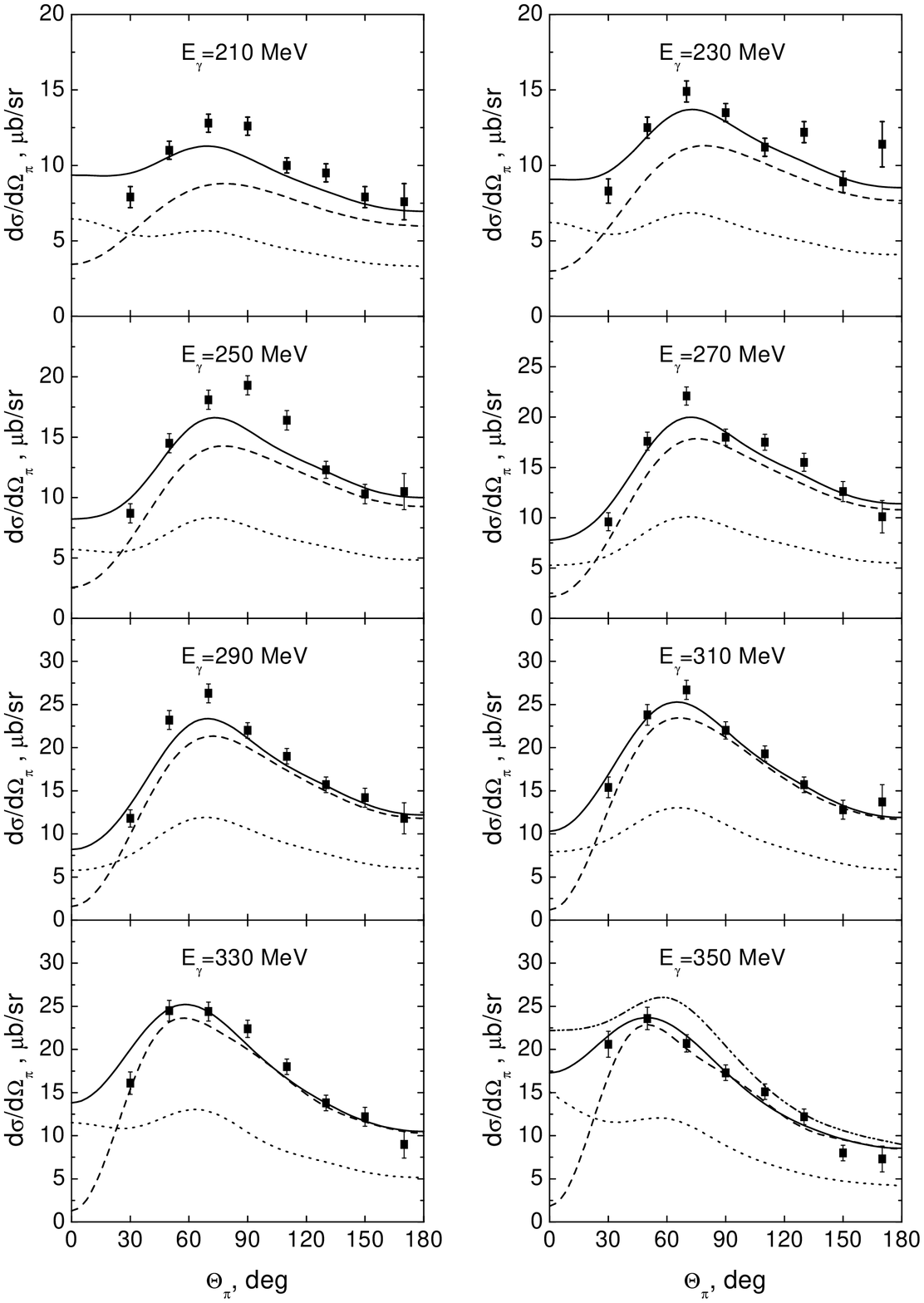}}
\caption{ 
Differential cross sections for the reaction 
$d(\gamma,\pi^-)pp$ in the lab frame. 
The dotted curves are contributions of one of the pole diagrams
in Fig.~\ref{fig1}. Successive addition of the second pole 
diagram and FSI leads to dashed and full curves, respectively.
At 350 MeV the results from Ref.~{\protect \cite{laget81}} are shown
as a dash-double-dotted curve.
Data are from Ref.~{\protect \cite{benz73}}. 
}
\label{fig4} 
\end{figure}

\setcounter{figure}{3}
\begin{figure}
\epsfxsize=0.85\textwidth
\centerline{\epsfbox[50 200 550 810]{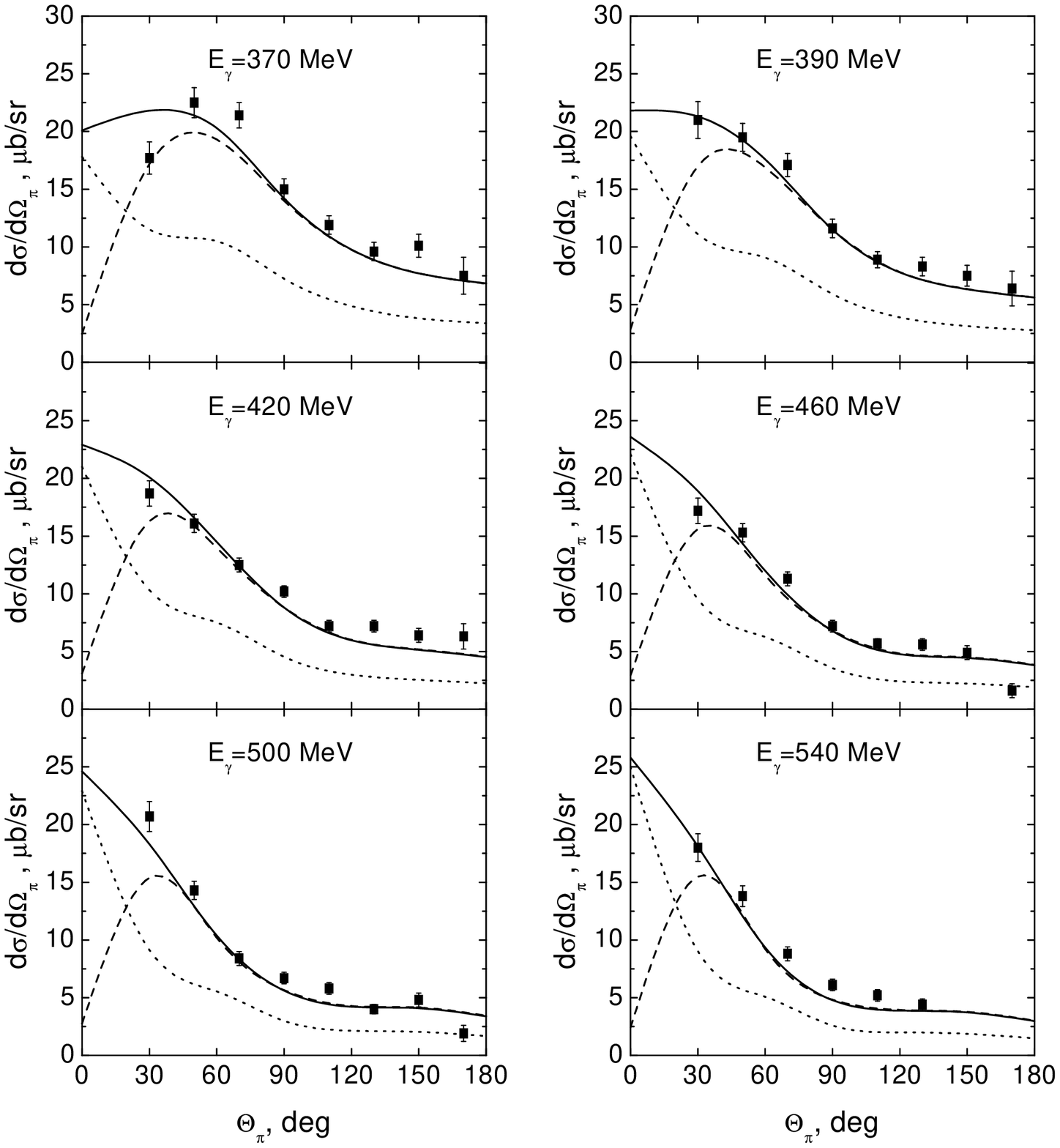}}
\caption{ Continued.
}
\label{fig4_2}
\end{figure}

\begin{figure}
\epsfxsize=0.8\textwidth
\centerline{\epsfbox[40 330 450 810]{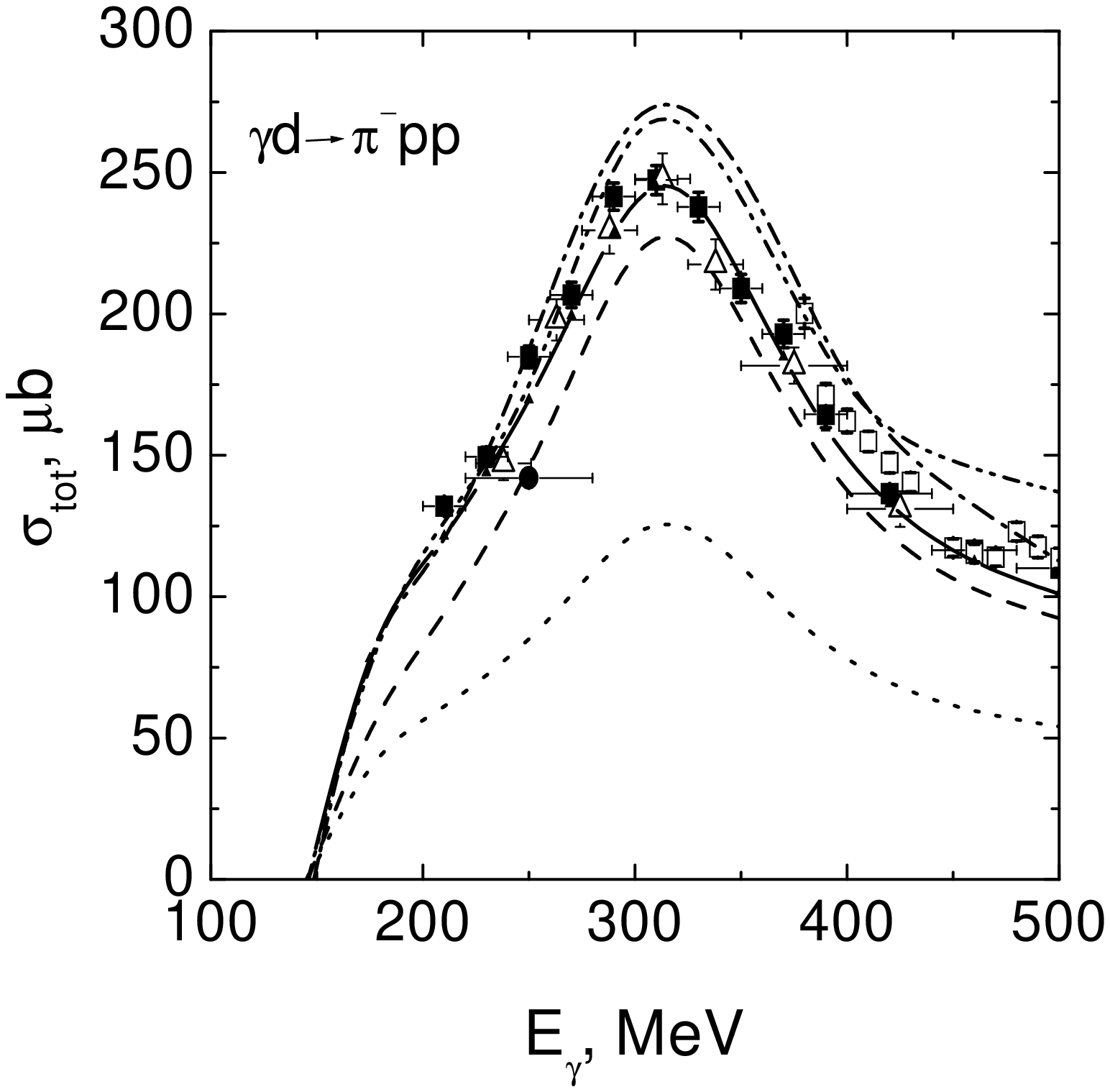}}
\caption{
Total cross section for the reaction 
$d(\gamma,\pi^-) pp$. Meaning of the curves as in Fig.~\ref{fig4}.
In addition the results from Refs.~{\protect \cite{laget81}} and 
{\protect \cite{schmidt96}} are shown  in dash-double-dotted and 
dash-dotted curves, respectively.  Data are from Refs.~{\protect 
\cite{benz73}} (solid boxes), {\protect \cite{chief75}} (empty 
triangles), {\protect \cite{asai90}} (empty boxes), and {\protect 
\cite{quraan98}} (solid circle).  
} 
\label{fig5} 
\end{figure}

\begin{figure}
\epsfxsize=0.85\textwidth
\centerline{\epsfbox[50 370 550 810]{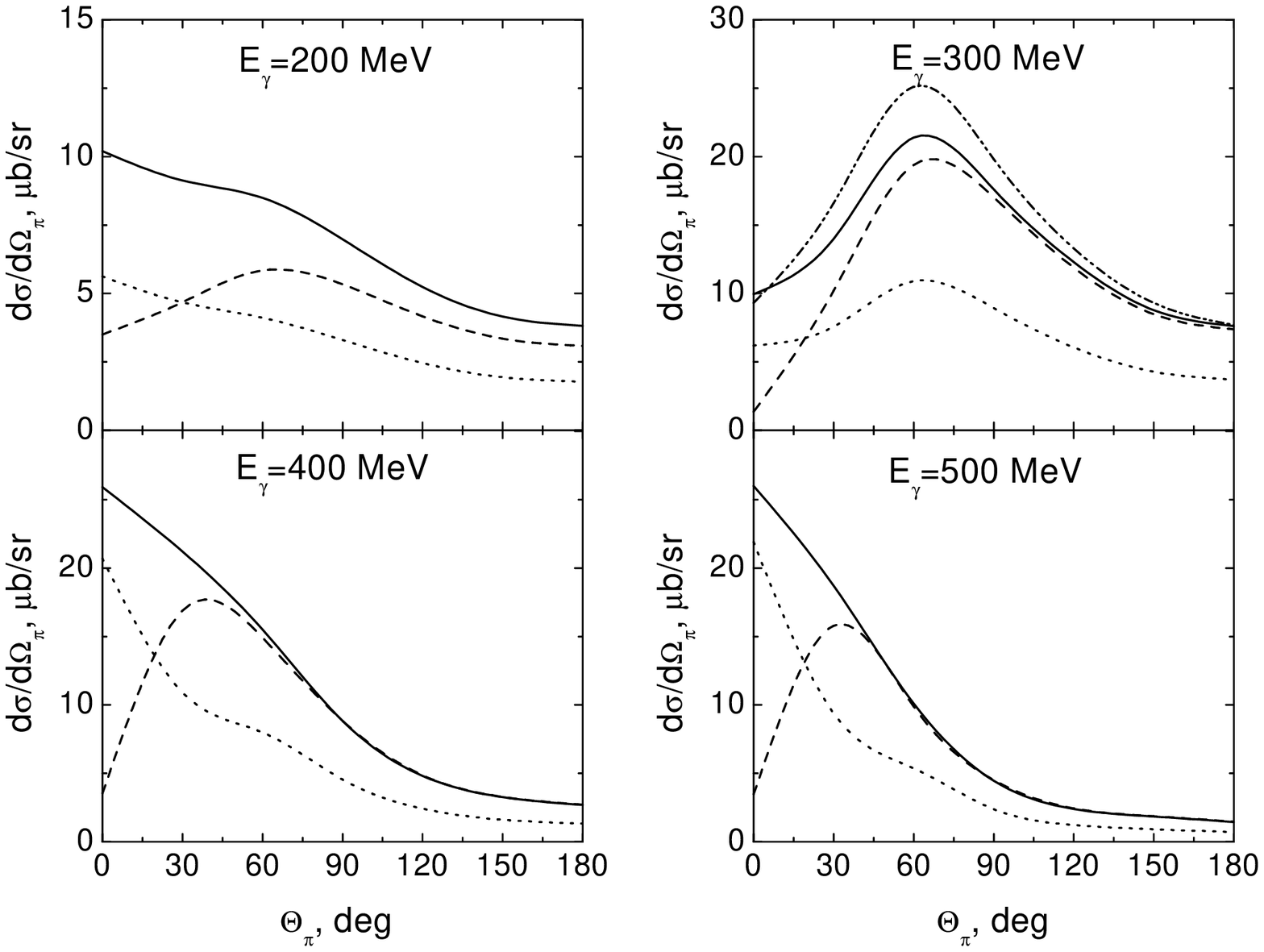}}
\caption{ 
Differential cross sections for the reaction 
$d(\gamma,\pi^+)nn$ in the lab frame at four energies. 
Meaning of the curves as in Fig.~\ref{fig4}.
}
\label{fig6} 
\end{figure}

\begin{figure}
\epsfxsize=0.8\textwidth
\centerline{\epsfbox[40 330 450 810]{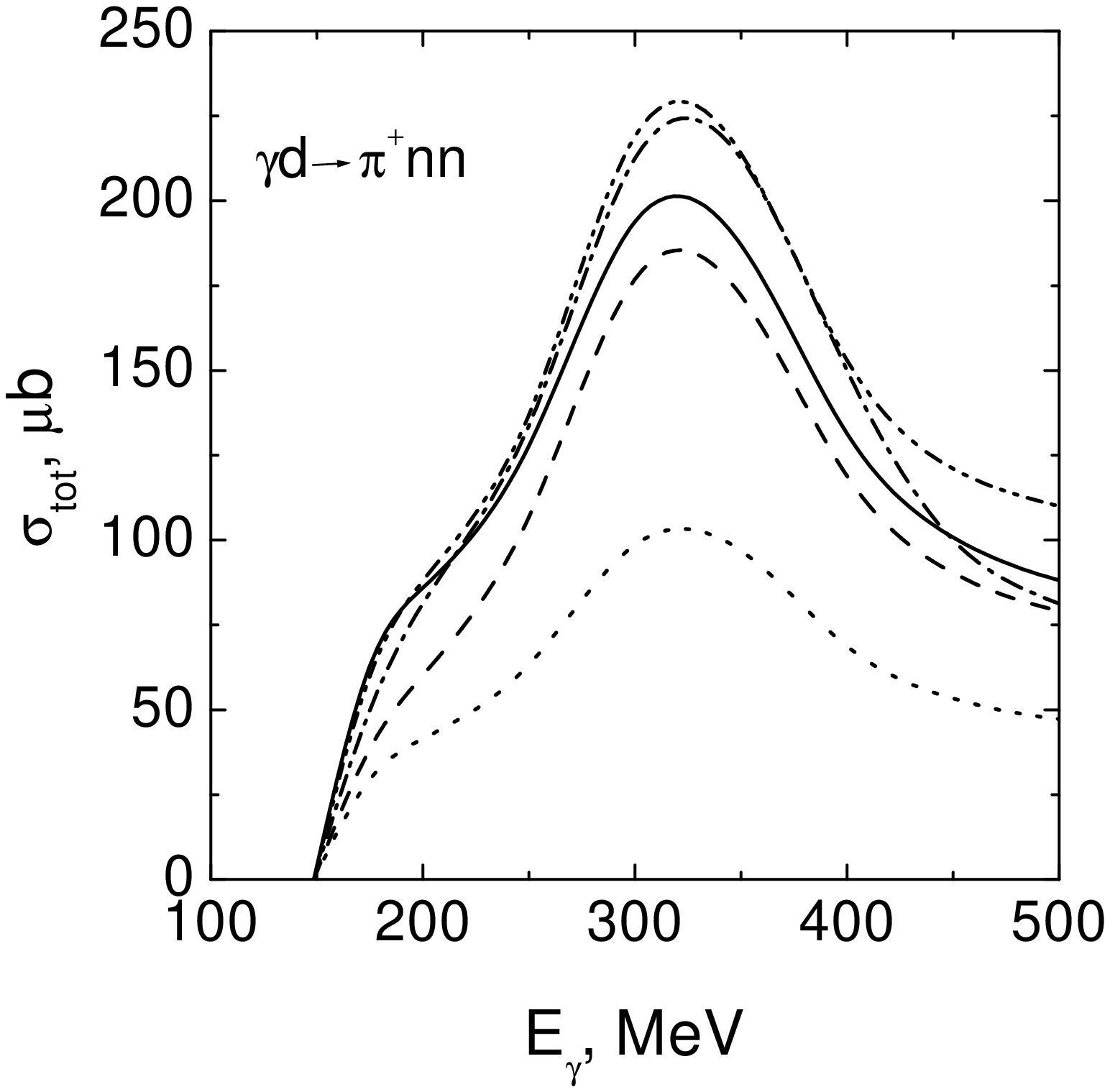}}
\caption{
Total cross section for the reaction 
$d(\gamma,\pi^+)nn$. Meaning of the curves as in Fig.~\ref{fig5}.
}
\label{fig7}
\end{figure}

\begin{figure}
\epsfxsize=0.8\textwidth
\centerline{\epsfbox[40 330 450 810]{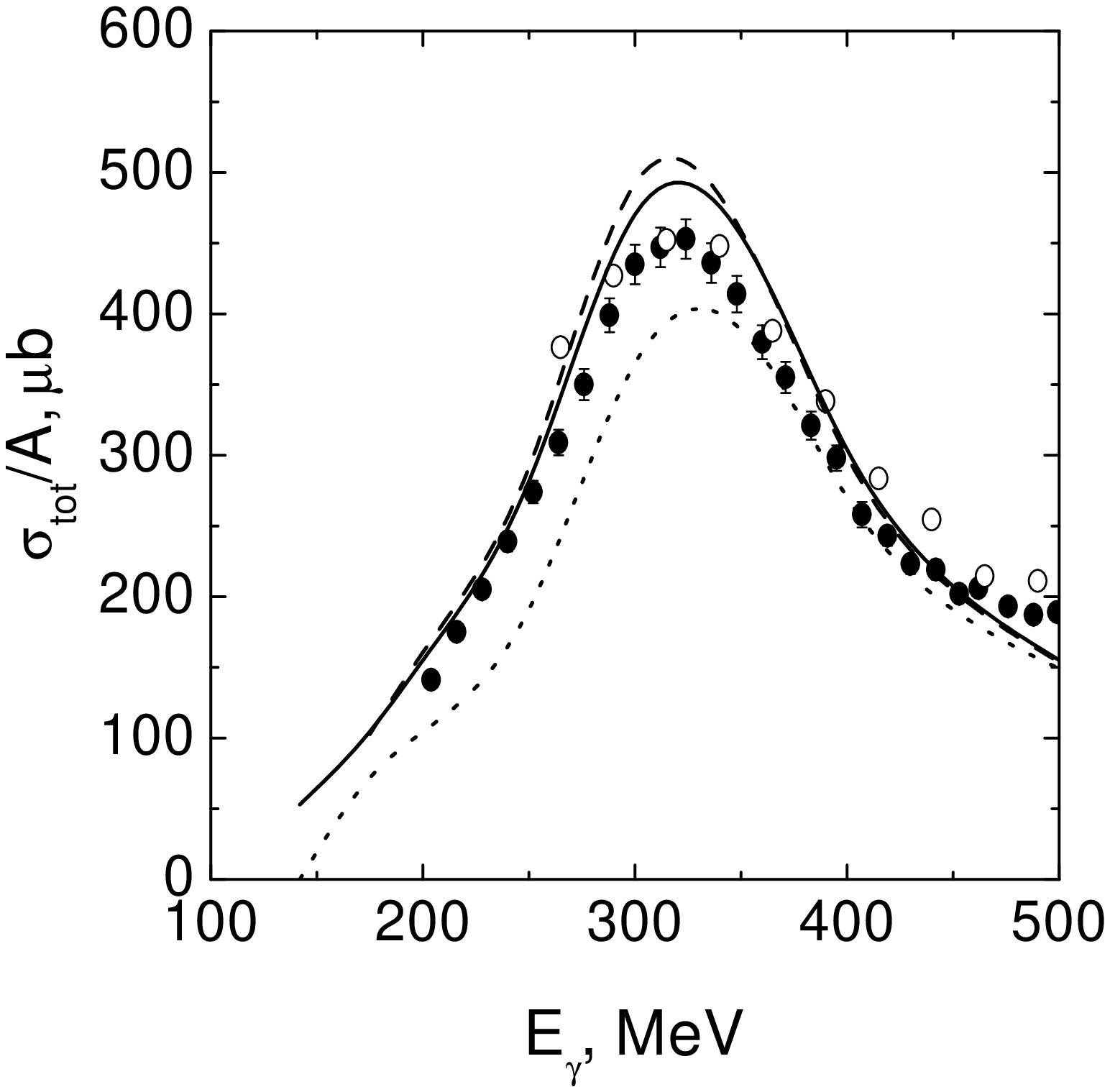}}
\caption{
Total photoabsorption cross section per nucleon for the deuteron from 
150 to 500 MeV.  Contributions from the reactions $\gamma d\to \pi 
NN$ are shown as dotted curve. Contributions from the reactions 
$\gamma d\to\pi^0 d$ and $\gamma d\to np$ are included in the full 
curve.  
Results with the MAID multipoles are shown in dashed curve.
Data are from Refs.~{\protect \cite{amstrong72}} (empty circles)
and {\protect \cite{maccornick96}} (solid circles).
}
\label{fig8}
\end{figure}

\end{document}